# Strain effect engineered in α-Al$_2$O$_3$/monolayer MoS$_2$ interface by first principle calculations

Sheng Yu [a,*], Shunjie Ran [b], Hao Zhu [a,c], Kwesi Eshun [a], Chen Shi [a], Kai Jiang [a], Qiliang Li [a]

[a] Department of Electrical and Computer Engineering, George Mason University, Fairfax, VA 22030, USA

[b] Department of Physics and Astronomy, George Mason University, Fairfax, VA 22030, USA

[c] State Key Laboratory of ASIC and System, School of Microelectronics, Fudan University, Shanghai 200433, China

**Abstract**

With the advances in low dimensional transition metal dichalcolgenides (TMDCs) based metal–oxide–semiconductor field-effect transistor (MOSFET), the interface between semiconductors and dielectrics has received considerable attention due to its dramatic effects on the morphology and charge-transport of semiconductors. In this study, first principle calculations were utilized to investigate the strain effect induced by the interface between α-Al$_2$O$_3$ (0001)/monolayer MoS$_2$. The results indicate that Al$_2$O$_3$ in 1.3nm-thickness can apply the strain of 0.3% on MoS$_2$ monolayer. The strain effect monotonically increases with the larger thickness of the dielectric layer. Also, the study on temperature effect indicates the monotonic lattice expansion induced by the higher temperature. Our study proposes that the dielectric engineering can be an effective tool for strain effect in the nanotechnology.

## I. INTRODUCTION

The semiconductor electronics industry is now facing serious challenges in promoting further scaling and reduction of power consumption. It is responding to these challenges by developing alternative semiconductor materials that can overcome the limitations of Si. [1-5] Low-dimensional nanostructures are attractive as a channel in future ultimately scaled transistors since their atomic scale thickness offers efficient electrostatic control, making them immune to short channel effects. [6,7] Graphene was known to have remarkable electrical and mechanical properties, [7] including high carrier mobility high thermal conductance, [8] and excellent stiffness. [9] However, the absence of intrinsic energy bandgap obstructs its further application in logic and memory devices which require high on-off ratio and low off-state current. [10,11] MoS$_2$, as a typical member of TMDCs, has the intrinsic direct bandgap (Eg ~ 1.8 eV).[12,13] This makes it promising for application in future field-effect transistors (FETs) with excellent current on/off ratio (>$10^8$).[13,14] Logic circuits and amplifiers based on monolayer MoS2 have also been demonstrated recently,[15,16] as well as saturation and high breakdown currents.[17] Also, the broken inversion symmetry induced by monolayer TMDCs endows MoS2 with natural piezoelectricity, rendering it a promising candidate for applications in mechano-electric converter.[18,19] The ultrasensitive monolayer MoS2 photodetector was further demonstrated with high photoresponsivity.[20]

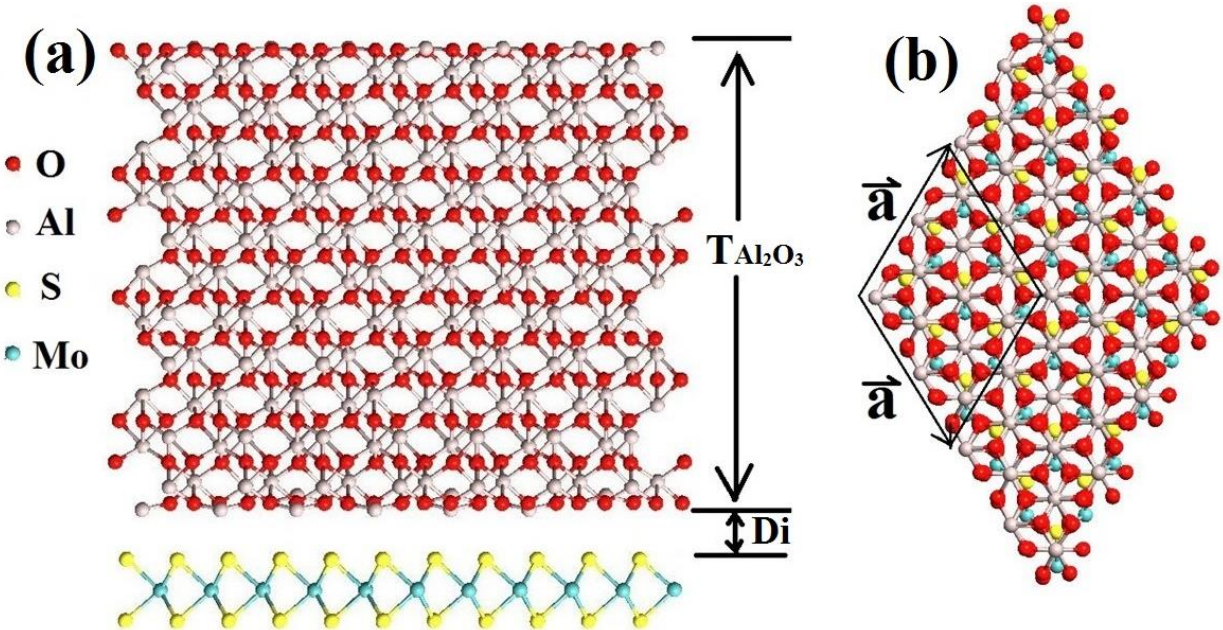

Figure 1. Schematic of 1.3nm-layer $Al_2O_3$ (0001) and $MoS_2$ monolayer interface $T_{Al_2O_3}$ represents the thickness of the dielectric layer, while Di represents the distance between $Al_2O_3$ and $MoS_2$.

Although possessing the promising multi-applications of MoS2 layered structures, it is still facing challenges in industrially applicable devices. MoS2 monolayer has low carrier mobility, about several tens of $cm^2$/Vs, limiting its application in high-performance FETs.[21,22] Also, the carrier transport in these 2D monolayers is heavily affected by the scattering of the acoustic phonon via intra- and inter-valley deformation potential coupling.[22,23] This further reduces the electrical conductance at room temperature (RT). In addition, the large variation in electrical properties induced by doping and strain in MoS2 monolayer could affect its applications in nanoelectronics.[12] Strain effect was proposed as to effectively improve the carrier mobility of MoS$_2$ monolayer.[24] Since MoS$_2$ can endure large strains before breaking, the modification of the MoS$_2$ band gap by applying strain becomes an important strategy to enhance the performance of nano-devices made from MoS$_2$.[25] In this study based on first principle calculations, we thoroughly explore the strain effect induced by the interface between α-Al$_2$O$_3$ (0001)/MoS$_2$ monolayer with various thickness of the dielectric layer. Our study indicates that the tensile strain of 0.3% on MoS$_2$ monolayer can be induced from the interface between 1.3nm-Al$_2$O$_3$/MoS$_2$ monolayer. The strain monotonically increases up to 0.62% until the dielectric thickness increases to 5.2nm. The investigation on the bandstructure indicates that all the interface exhibit an indirect bandgap with valence band maximum (VBM) located at Γ point and conduction band minimum (CBM) at M point in Brillouin Zone. Also, the higher temperature can induce an expanded lattice. Our study demonstrates the strain engineering of the interface on the TMDCs semiconductor, which significantly improves the carrier mobility and enhances the performances in MOSFET.

## II. METHODOLOGY

In this study, first principle calculations were carried out by using the Virtual Nanolab Atomistix ToolKit (ATK) package with the density functional theory (DFT).[26] The localized density approximation (LDA) exchange correlation with a double zeta polarized (DZP) basis was used with a mesh cut-off energy of 150 Ry.[27] All the atomic positions and lattice parameters were optimized by using the generalized gradient approximations (GGA) with the maximum Hellmann-Feynman forces of 0.05 eV/Å, which is sufficient to obtain relaxed structures.[28] The Pulay-mixer algorithm was employed as iteration control parameter with tolerance value of $10^5$.[29] The maximum number of fully self-consistent field (SCF) iteration steps was set to 1000.[27] The electronic temperature is set to 300 K for our simulations before considering the temperature effect. The periodic boundary condition was employed along all the three directions in the hexagonal lattice.[30] The distance between neighboring interface is set to be 30Å to minimize the interaction between them. The self-consistent field calculations were checked strictly to guarantee fully converging within the iteration steps.[27]

Figure 1 demonstrates the schematic of interface between 1.3nm-Al$_2$O$_3$ （0001）/ MoS$_2$ monolayer. T$_{Al_2O_3}$ denotes the thickness of the dielectric layer while D$_i$ indicates the separation distance between Al$_2$O$_3$/MoS$_2$, which is calibrated by the nearest vertical distance between Oxygen and Sulfur. The unit cell of the crystal with the hexagonal lattice is enclosed by the parallelogram shown in the Fig. 1(a). This unit cell stems from merging together the unit cell of $Al_2O_3$ in repetition of 2×2 and $MoS_2$ monolayer by 3×3.[24,31] In this study, we investigate the strain effect from this interface by following steps: (a) Set the thickness of $Al_2O_3$ to 1.3nm and investigate the intrinsic separation between $Al_2O_3$/ $MoS_2$. (b) Investigate the evolution of the total energy of the unit cell vs. lattice parameter ($L_c$) and figure out the intrinsic $L_c$ corresponding to the minimum total energy. Then calculate the strain effect by comparison of $L_c$ with the lattice parameter of $MoS_2$ monolayer ($L_0$): $\varepsilon = (L_c - 3L_0)/3L_0$. (c) Investigate the strain effect by varying the thickness of dielectric layer to 1.3nm, 2.6nm, 3.9nm and 5.2nm.

## III. RESULTS AND DISCUSSION

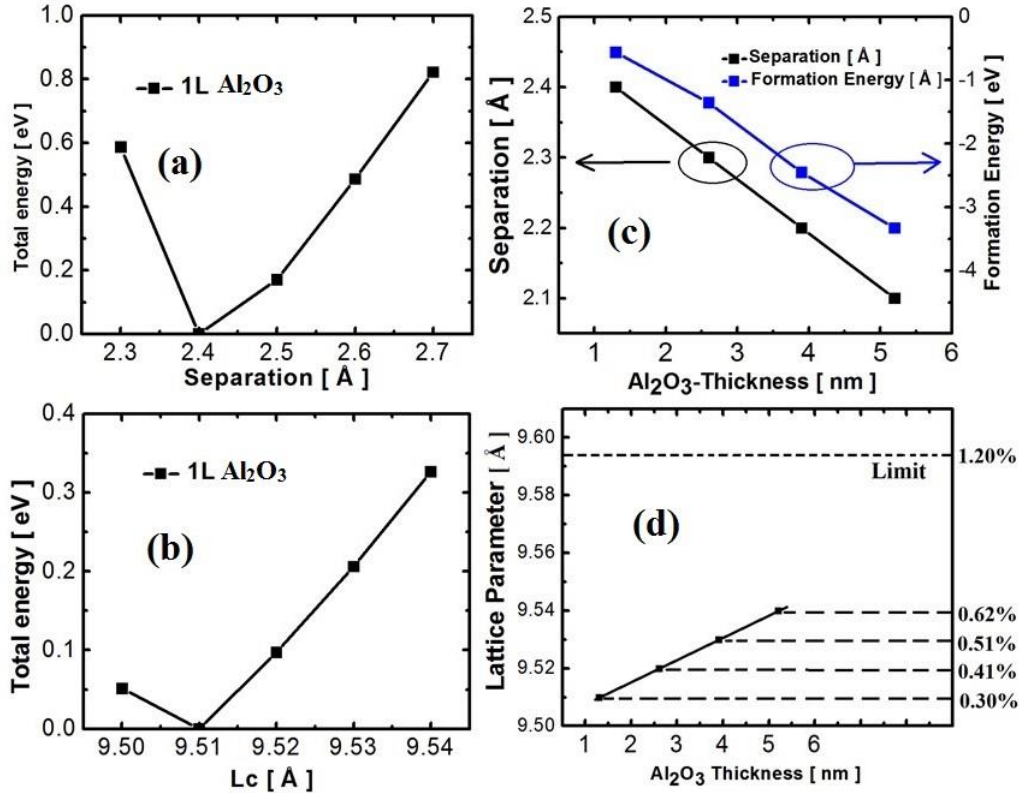

Figure 2. The total energy of the unit cell vs. (a) separation between $MoS_2$ and $Al_2O_3$ (b) lattice parameter ($L_c$)

Figure 2 explores the intrinsic lattice parameter of this interface. Fig. 2(a) demonstrates the evolution of the total energy of the unit cell with the separation distance between $MoS_2$ and $Al_2O_3$. It exhibits an intrinsic Di of 2.4Å if the thickness of dielectric layer is fixed to 1.3nm. Fig. 2(b) indicates the evolution of the total energy of the unit cell with $L_c$. As shown the intrinsic $L_c$ is 9.51Å in case of 1.3nm-thickness of the dielectric layer. The calculated strain effect from this interface on MoS$_2$ monolayer is 0.3% by the mathematical expression: $\varepsilon =(L_c-3L_0)/3L_0$. Fig. 2(c) explores the structural stability of the interface. The formation energy for each structure is denoted by the mathematical expression: $E_F=E_{MoS_2} + E_{Al_2O_3} - E_i$. $E_i$ denotes the total energy of the interface, $E_{MoS_2}$ and $E_{Al_2O_3}$ represent the structual energy of MoS$_2$ monolayer and Al$_2$O$_3$ before merging, respectively.[32] The total energy consists of the exchange-correlation energy, the kinetic energy, and the electrostatic energy.[33] As shown the formation energy monotonically decreases with the larger thickness of the dielectric layer, demonstrating the enhanced structural stability. We also investigate the effect of the thickness of Al$_2$O$_3$ on the intrinsic separation distance between the dielectric layer and MoS$_2$ monolayer. Fig. 2(c) indicates that the larger thickness of the dielectric layer leads to the smaller separation, where it possesses the separation of 2.1Å if the thickness of dielectric layer is grown to 5.2nm. We further explore the effect of the thickness of the dielectric layer on the strain effect. As demonstrated in Fig. 2(d) the larger thickness of the dielectric layer monotonically boosts the strain effect from 0.30% to 0.62% when the thickness increases from 1.3nm to 5.2nm. It is noteworthy that there is the strain limit of 1.2% when Al$_2$O$_3$ thickness increases to infinite, where the lattice parameter of the interface extremely approaches to that of the bulk Al$_2$O$_3$.

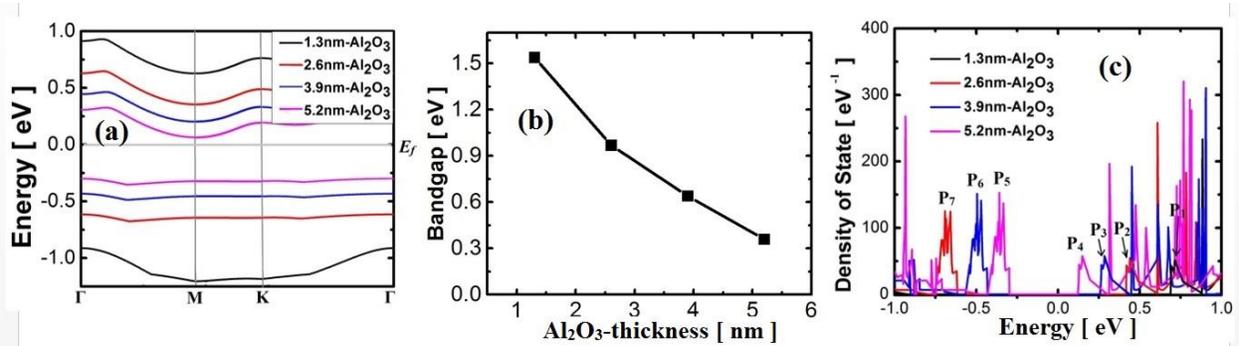

Figure 3. The effect of Al$_2$O$_3$-thickness on (a) bandstructure; (b) bandgap; (c) density of states. The Fermi level is pinned to 0eV.

The effect of the thickness of the dielectric layer on the electronic properties of the interface is further investigated. Fig. 3(a) shows the bandstructure of the interface for the thickness of $Al_2O_3$ at 1.3nm, 2.6nm, 3.9nm and 5.2nm, respectively. VBM and CBM are remaining at Γ point and M point in the Brillouin Zone respectively for all the structures. Fig. 3(b) demonstrates the reduced indirect bandgap with continuously increased thickness of the dielectric layer. The interface with 5.2nm-$Al_2O_3$ exhibits the small bandgap of 0.36eV. Fig. 3(c) displays the density of states (DOS) of the interface with different thickness of the dielectric layer. As shown the DOS around CBM ($P_1$, $P_2$, $P_3$, $P_4$) move toward the Fermi Level with the increased thickness of $Al_2O_3$. All these peaks arise from the atomic shells of $MoS_2$. However, the DOS around VBM ($P_5$, $P_6$, $P_7$) originate from the atoms of $Al_2O_3$. They also approach the Fermi Level with the larger thickness of $Al_2O_3$.

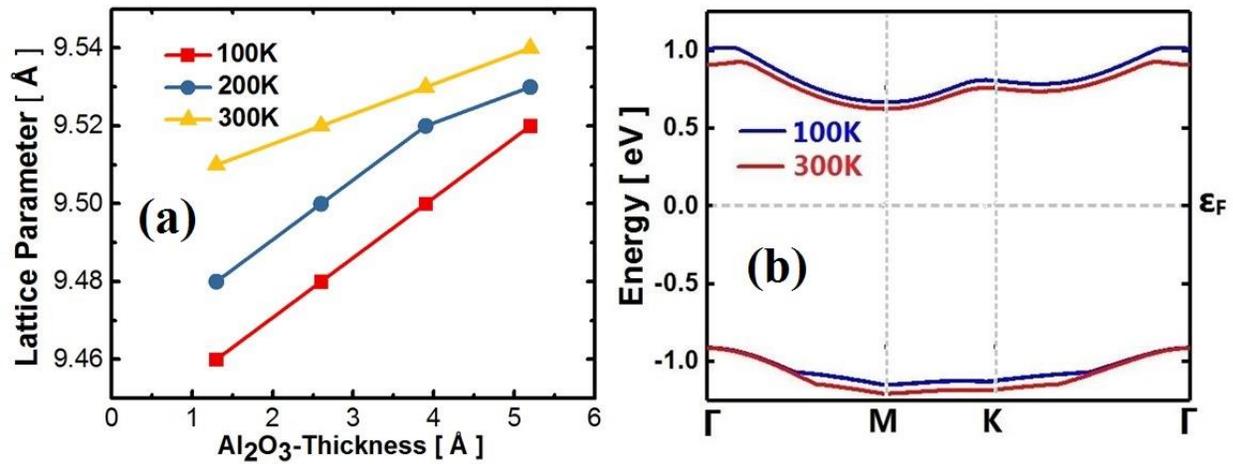

Figure 4. (a)The temperature effect on lattice parameter ($L_c$); (b) The comparison of the bandstructure

Lastly, we consider the temperature effect on the lattice expansion and the bandstrcucture. Fig. 4(a) exhibits the evolution of the intrinsic lattice parameter of the interface with the various thickness of $Al_2O_3$ under different temperatures of 100K, 200K and 300K, respectively. The lattice parameter increases monotonically with the higher temperature.[34] We notice that the interface with smaller size is affected more significantly by the higher temperature. $L_c$ of 1.3nm- $Al_2O_3$/$MoS_2$ increases from 9.46Å to 9.51Å by the lattice expansion of 0.53% when the temperature is increased from 100K to 300K, whereas it only increases by 0.21% in the case of 5.3nm-$Al_2O_3$/$MoS_2$. This arises from the larger sensitivity in the thermal expansion of $MoS_2$ than $Al_2O_3$ induced by the higher temperature.[35,36] The bandstructure of 1.3nm-$Al_2O_3$/$MoS_2$ under 100K and 300K is compared in Fig. 4(b). We note that due to the lattice expansion, the bandgap reduces slightly from 1.58eV to 1.54eV at the increased temperature from 100K to 300K. VBM and CBM remain at the same position with the high symmetry in the Brillouin Zone. The reduced bandgap induced by the higher temperature can lead to the redshift exciton emission in the optical detection.

## IV. CONCLUSION

In conclusion, assisted by the first principle calculation we comprehensively explore the strain effect induced by $Al_2O_3$/$MoS_2$ monolayer interface with the various thickness of dielectric material. Our study indicated 0.3% strain on $MoS_2$ monolayer is induced from interface between 1.3nm-$Al_2O_3$ (0001)/ $MoS_2$ monolayer. The strain effect monotonically increase with the larger thickness of the dielectric layer, up to 0.62% with 5.2nm-$Al_2O_3$. The investigation on the bandstructure indicates that all the interface exhibit an indirect bandgap with VBM located at $\Gamma$ point and CBM at M point in Brillouin Zone. Also, the lattice expansion and the reduced bandgap are observed by the higher temperature. Our study proposes that the dielectric engineering can be an effective candidate for applying strain effect on channel semiconductors in MOSFET, which has promising applications in industrial and academic low-dimensional semiconductors.

## ACKNOWLEDGEMENTS

This work was supported in part by the U.S. NSF Grant ECCS-1407807 and in part by Virginia Microelectronics Consortium research grant. The authors declare no competing financial interest.